\documentclass[letterpaper,10pt,conference]{ieeeconf}
\IEEEoverridecommandlockouts
\usepackage{silence}
\usepackage{amsmath,amssymb,amsfonts,mathtools,mathrsfs,bm}
\usepackage{graphicx,booktabs,cite}
\usepackage{subcaption}
\usepackage[ruled,vlined]{algorithm2e}
\SetKw{AlgReturn}{return}
\SetKw{AlgBreak}{break}
\SetAlCapFnt{\small}
\SetAlCapNameFnt{\small}

\usepackage{hyperref}

\usepackage{tikz}
\usetikzlibrary{arrows.meta, positioning, calc, fit, backgrounds}

\usetikzlibrary{decorations.markings}
\definecolor{figOwn}{HTML}{1F5FA8}   
\definecolor{figComm}{HTML}{2A7F62}  
\definecolor{figSea}{HTML}{9CC3DD}   

\newcommand{\R}{\mathbb R}

\newcommand{\X}{\mathscr X}
\newcommand{\Y}{\mathscr Y}
\newcommand{\W}{\mathscr W}
\newcommand{\Z}{\mathscr Z}
\newcommand{\Ss}{\mathscr S}
\newcommand{\Hs}{\mathcal H}

\newcommand{\col}{\operatorname{col}}

\newcommand{\He}{\operatorname{He}}

\newtheorem{assumption}{Assumption}

\newtheorem{theorem}{Theorem}

\newtheorem{remark}{Remark}

\allowdisplaybreaks
\begin{document}
\title{Distributed Output-Feedback Control for Linear Multichannel Systems}
\author{Ruixuan~Zhao, Guitao~Yang, Boli~Chen, Thomas~Parisini
\thanks{R. Zhao and B. Chen are with the Department of Electronic and Electrical Engineering and with the Dynamic Systems Lab, University College London, London, UK
{\tt\small (ruixuan.zhao.22@ucl.ac.uk; boli.chen@ucl.ac.uk).}}
\thanks{G. Yang is with the Department of Computer Science, City St. George’s, University of London, London, UK \tt\small(guitao.yang@city.ac.uk).}
\thanks{T. Parisini is with the Department of Electrical and Electronic Engineering, Imperial College London, London. T. Parisini is also with the Department of Electronic Systems, Aalborg University, Denmark, and with the Department of Engineering and Architecture, University of Trieste, Italy \tt\small(t.parisini@imperial.ac.uk).}
}
\maketitle
\begin{abstract}
This paper investigates distributed output-feedback stabilization of linear multichannel systems in which agents exchange state estimates without disclosing local control inputs or feedback gains. Remote control actions are treated as unknown inputs and geometrically decoupled from stable quotient-error dynamics. The proposed framework accommodates nontrivial intersections of local loss subspaces by retaining the coupling between the plant and common estimation errors. A Hurwitz condition on the coupled dynamics, together with a sufficient consensus-gain bound, guarantees exponential closed-loop convergence for every fixed feedback selection in a compact family. For polytopic feedback families, a quotient-preserving parameterization of output injections enables joint synthesis of nominal feedback and observer gains via alternating linear matrix inequalities, followed by a convex semidefinite program certifying a common consensus gain. A numerical example demonstrates the effectiveness of the proposed design.
\end{abstract}

\section{Introduction}
Many engineering systems are actuated and sensed by spatially distributed agents acting on a shared plant, as in cooperative multi-agent systems~\cite{olfati2007consensus}, vehicle platoons~\cite{qiang:tits23}, and microgrids~\cite{dorfler2016hierarchy}. A linear multichannel model assigns each agent a local input--output channel through which it measures a partial output and applies a control input. Here, ``multichannel'' describes the plant structure, whereas ``centralized'', ``decentralized'', and ``distributed'' describe the information available during operation. A centralized controller processes all channels together, decentralized controllers use only their own channels, and distributed controllers additionally communicate with neighboring agents. A separate issue is which model parameters, control inputs, and feedback gains each agent must know or disclose.

Early work on decentralized control identified fixed modes as a fundamental obstruction. It was shown in~\cite{wang1973decentralized} that stabilization by finite-dimensional linear time-invariant (LTI) decentralized controllers is possible if and only if all decentralized fixed modes are stable. Conditions for decentralized stabilization and spectrum assignment based on channel interactions and complementary subsystems were developed in~\cite{corfmat1976decentralized}. Fixed modes can also arise from structural constraints rather than exact parameter coincidences. Their generic absence is characterized through structural completeness, including systems with dependent parameters~\cite{liu2021structural}. These limitations motivate inter-agent communication.

The question is then what the agents should communicate. The results in~\cite{liu2020does,liu2023distributed} show that sharing neighboring measurements alone need not eliminate fixed modes, even when the plant is jointly controllable and observable and the communication graph is strongly connected. They also establish conditions on the dimensions of shared controller states that remove this obstruction. Exchanging control inputs or feedback gains may be undesirable when local actions and policies are not to be disclosed. A natural alternative is to exchange local state estimates, which serve as communicated controller states and support estimate-based feedback at each agent.

\begin{figure}[t]
  \centering
\begin{tikzpicture}[font=\footnotesize, >={Stealth[length=1.7mm]},
  agent/.style={circle, draw=black!55, fill=black!8, line width=0.7pt, minimum size=0.62cm, inner sep=0pt},
  me/.style={agent, draw=figOwn, fill=figOwn!18, line width=1.1pt},
  port/.style={rounded corners=1pt, minimum width=0.34cm, minimum height=0.13cm, inner sep=0pt}]
\shade[top color=black!3, bottom color=black!15, draw=black!70, line width=0.9pt] (0,0) ellipse (2.2 and 1.0);
\node at (0,0.12) {$\dot x = Ax + \sum_{j=1}^{N} B_j u_j$};
\node[black!55, font=\scriptsize] at (0,-0.3) {shared plant};
\def\chan#1#2#3#4{\coordinate (p#1) at #2; \coordinate (a#1) at #4; \def\ang{#3}%
  \node[port, rotate=\ang+90, fill=black!45] at (p#1) {};}
\chan{N}{(-2.2,0)}{180}{(-3.5,0)}
\chan{1}{(-0.93,0.906)}{102}{(-1.13,1.84)}
\chan{j}{(2.067,0.342)}{39}{(3.08,1.16)}
\node[port, rotate=-75+90, fill=figOwn] at (1.1,-0.866) {};
\coordinate (pi) at (1.1,-0.866); \coordinate (ai) at (1.34,-1.79);
\foreach \n/\a/\uside/\yside in {N/180/above/below, 1/102/right/left, j/39/below right/above left}{
  \draw[->, black!40, line width=0.6pt] ($(p\n)+(\a+90:0.09)$) -- node[\yside, inner sep=1pt, black!50, font=\scriptsize]{$y_{\n}$} ($(a\n)+(\a+90:0.09)+(\a:-0.31)$);
  \draw[<-, black!45, line width=1.0pt, densely dotted] ($(p\n)+(\a-90:0.09)$) -- node[\uside, inner sep=1pt, black!55, font=\scriptsize]{$u_{\n}$} ($(a\n)+(\a-90:0.09)+(\a:-0.31)$);}
\draw[->, figOwn, line width=0.8pt] ($(pi)+(-75-90:0.09)$) -- node[left, inner sep=1pt]{$y_i$} ($(ai)+(-75-90:0.09)+(-75:-0.31)$);
\draw[<-, figOwn, line width=1.1pt] ($(pi)+(-75+90:0.09)$) -- node[right, inner sep=1pt]{$u_i$} ($(ai)+(-75+90:0.09)+(-75:-0.31)$);
\node[figOwn, font=\scriptsize, anchor=south] at ($(pi)+(0.02,0.08)$) {$B_i,C_i$};
\node[agent] at (aN) {$N$};
\node[agent] at (a1) {$1$};
\node[agent] at (aj) {$j$};
\node[me]    at (ai) {$i$};
\node[figOwn, anchor=west, align=left, inner sep=1pt] at ($(ai)+(0.38,-0.10)$) {\textbf{agent $i$}\\ $u_i = R_i H_i\hat x_i$};
\begin{scope}[figComm, line width=0.7pt, dash pattern=on 2.5pt off 1.5pt]
  \draw ($(aN)+(0,0.31)$) to[out=90,in=200] ($(a1)+(-0.3,-0.06)$);
  \draw ($(a1)+(0.3,0.06)$) to[out=10,in=110] node[pos=0.45, above, inner sep=1.5pt, solid]{$\hat x$} ($(aj)+(0,0.31)$);
  \draw ($(aj)+(0.22,-0.22)$) to[out=-70,in=25] node[pos=0.35, right, inner sep=2pt, solid]{$\hat x_j,\ \hat x_i$} ($(ai)+(0.28,0.14)$);
\end{scope}
\node[draw=figOwn, fill=white, rounded corners=2pt, anchor=north east, inner sep=2.5pt, font=\scriptsize]
  at ($(ai)+(-0.45,0.25)$) {\begin{tabular}{@{}l@{\;\;}l@{}}
    \textcolor{figOwn}{\textbf{known}}       & $A,\ B_i,\ C_i,\ H_i$\\
    \textcolor{figComm}{\textbf{exchanged}}  & $\hat x_i,\ \hat x_j,\ (i,j)\in\mathcal E$\\
    \textcolor{black!55}{\textbf{unknown}}   & $u_j,\ y_j,\ H_j,\ j\neq i$
  \end{tabular}};
\end{tikzpicture}
  \caption{Distributed ouput-feedback control framework for linear multichannel systems. Each agent accesses the shared plant through its local input--output channel and exchanges only state estimates with its neighbors over a communication network (green dashed line). Remote control inputs (gray dotted arrows) are are unavailable to agent $i$ and are treated as unknown inputs by its local observer.
  }
  \vspace{-2ex}
  \label{fig:multichannel}
\end{figure}
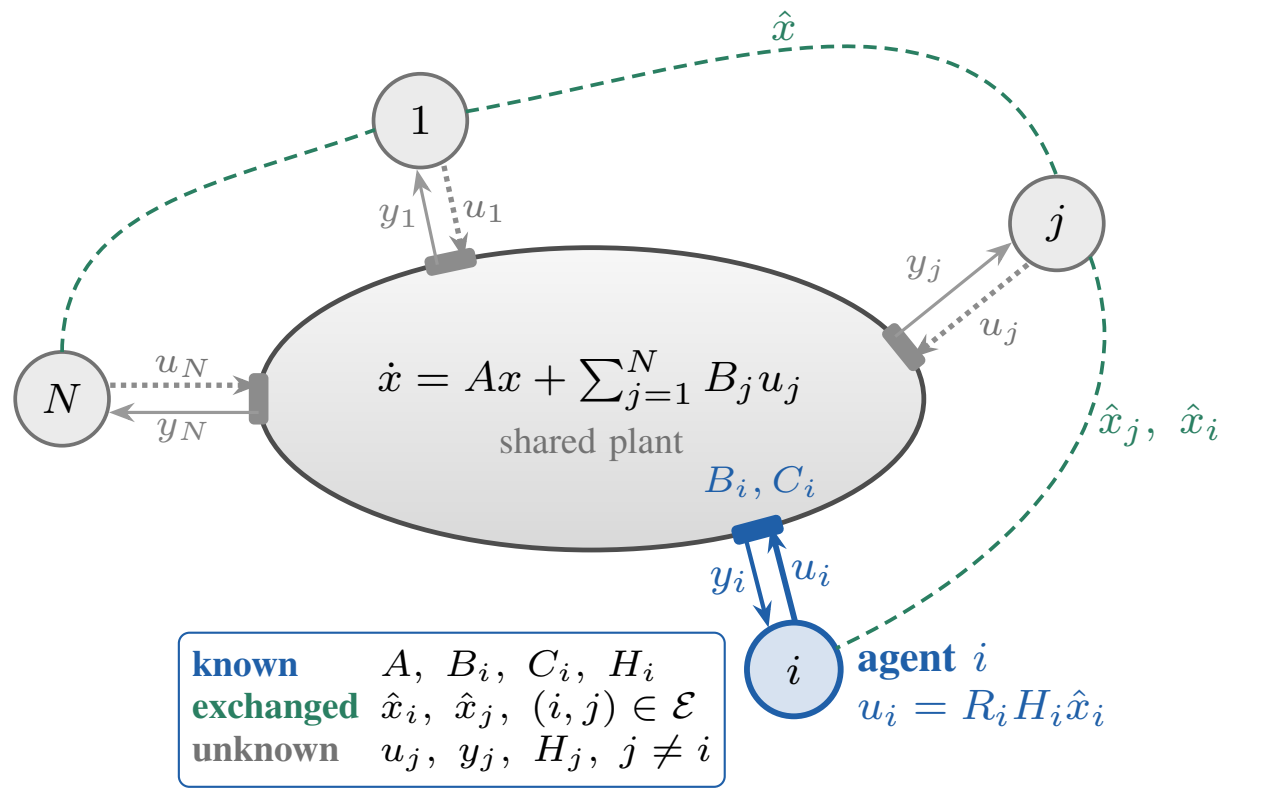

The main difficulty is that an agent with access only to its local input cannot directly reproduce the aggregate plant input in its observer. The input-cancellation mechanism used in centralized observer-based control is therefore generally unavailable. The observer-based constructions in~\cite{wang2020distributed,liu2023distributed} replace remote estimates by the local estimate through the substitution rule, but retain the aggregate feedback matrix in each observer. The discrete-time weak-separation result in~\cite{savas2022separation} retains this dependence. The scaled-local-input construction in~\cite{kim2024multichannel} avoids remote feedback gains and enables decentralized gain computation and plug-and-play operation. Its fixed-gain result requires Hurwitz state-feedback and aggregate observer dynamics. Under collective strong detectability, \cite{cao2025functional} combines local unknown-input decoupling with adaptive consensus to enable separate observer and controller design.

From the viewpoint of each agent, remote control actions are unknown inputs confined to known actuation subspaces. Geometric control theory~\cite{wonham1985,massoumnia1986geometric,basile1992} uses conditioned-invariant subspaces to decouple such inputs from appropriate estimation-error dynamics. Building on this approach, distributed unknown-input observers (DUIOs)~\cite{zhao:ecc25,zhao2025bridging,zhao2026duio} decouple locally unknown inputs from stable quotient-error dynamics and recover the remaining state information through communication. These designs address state estimation under exogenous unknown inputs. In feedback, however, the unknown inputs are generated by the local state estimates. When local loss subspaces intersect nontrivially, consensus cannot suppress the common-error directions. The resulting plant--common-error coupling motivates joint stabilization with observer parameters fixed across an admissible feedback family.

Building on this DUIO framework, this paper develops a linear distributed observer--controller scheme in which agents exchange only state estimates. First, for prescribed actuation subspaces, remote control actions are geometrically decoupled from stable local quotient-error dynamics. The online implementation uses local measurements, actual inputs, and neighboring estimates without exchanging control inputs or feedback gains. Second, the plant--common-error coupling is retained to accommodate nontrivial intersections of local loss subspaces, without separately requiring the selected feedback gains to stabilize the plant under full-state feedback. A reduced Hurwitz condition and a matrix-based consensus-gain bound guarantee exponential state and estimation-error convergence for every feedback selection in a compact family. Third, for polytopic feedback families with prescribed gain offsets, a quotient-preserving injection parameterization enables joint synthesis of nominal feedback and observer gains via alternating linear matrix inequalities (LMIs). A subsequent convex semidefinite program (SDP) certifies a common consensus gain over the resulting family.

Section~\ref{sec:problem} provides the notations, geometric preliminaries, and introduces the problem. Section~\ref{sec:dof_control} proposes the observer-controller design with closed-loop stability analysis. Section~\ref{sec:synthesis} gives the synthesis and gain optimization. Sections~\ref{sec:example} and~\ref{sec:conclusion} present the numerical results and conclusions, respectively.

\section{Preliminaries}\label{sec:problem}
\subsection{Notations and Geometric Elements}
Let $\mathbb R$ and $\mathbb C$ denote the real and complex fields, respectively. Set $\mathbb C_g=\{\lambda\in\mathbb C:\operatorname{Re}\lambda<0\}$ and $\mathbb C_b=\mathbb C\setminus\mathbb C_g$. Let $I_n$ denote the $n\times n$ identity matrix and $\otimes$ the Kronecker product. We use $\mathrm{col}(M_1,\ldots,M_n)$ and $\mathrm{diag}(M_1,\ldots,M_n)$ for block column stacking and block diagonal concatenation, respectively. $\kappa(M)$ denotes the spectrum of $M$. For a real symmetric matrix $M$, $\lambda_{\max}(M)$ denotes its maximum eigenvalue. $\He(M)=M+M^\top$. $\operatorname{co}\{\cdot\}$ denotes the convex hull of a set. The symbols $\|\cdot\|$ and $\|\cdot\|_2$ denote the Euclidean norm for vectors and the induced spectral norm for matrices. The vector $\mathbf 1_N$ has $N$ entries equal to one.

For $A:\X\to\X$, a subspace $\W$ is $A$-invariant if $A\W\subseteq\W$. In this case, $A|\X/\W$ denotes the induced map on the quotient space, satisfying $(A|\X/\W)P_\W=P_\W A$, where $P_\W$ is the canonical projection. The restriction of $A$ to $\W$ is denoted by $A|\W$. 
For $C:\X\to\Y$, a subspace $\W\subseteq\X$ is $(C,A)$-invariant if $(A+LC)\W\subseteq\W$ for some $L:\Y\to\X$. We denote the set of admissible injections by $\mathbf L(\W)$ and the family of $(C,A)$-invariant subspaces containing $\mathscr D$ by $\underline{\W}(C,A;\mathscr D)$. The symbol $\simeq$ denotes vector-space isomorphism. The notation $\mathrm{Mat}(\cdot)$ denotes the matrix representation in the specified bases. The symbol $\oplus$ indicates that the subspaces being added are independent. 
Following~\cite{wonham1985,massoumnia1986geometric}, all vector spaces are finite-dimensional over $\mathbb R$, and script letters denote subspaces.
Subspaces, invariance properties, and induced maps are defined coordinate-free.
For computation, we fix the standard bases, so that each linear map is identified with its matrix.

\subsection{Problem Formulation}
Consider the continuous-time linear multichannel system
\begin{equation}
 \dot x=Ax+\sum_{i=1}^N B_i u_i,\qquad y_i=C_i x,
 \quad i\in\mathbf{N},
 \label{eq:plant}
\end{equation}
where $x\in \X \simeq \R^n$, $u_i\in \mathscr U_i \simeq \R^{m_i}$, $y_i\in\mathscr Y_i\simeq\R^{p_i}$ are the system state, the local input, and the local measurement, respectively. 
The maps $A:\X\rightarrow\X$, $B_i:\mathscr U_i\rightarrow\X$, and $C_i:\X\rightarrow\mathscr Y_i$ are are identified with their matrices in the given bases. 
$\mathbf{N}=\{1,2,\cdots,N\}$ is the set of agents. 

The communication network over the agents is described by an undirected graph $\mathcal{G}=\{\mathbf{N},\mathcal{E},\mathcal{A}\}$, where $\mathcal{E}\subseteq\mathbf{N}\times\mathbf{N}$ is the set of communication links and $\mathcal{A}=[a_{ij}]\in\mathbb{R}^{N\times N}$ is the adjacency matrix, with $a_{ij}=a_{ji}=1$ if $(i,j)\in\mathcal{E}$ and $a_{ij}=0$ otherwise ($a_{ii}=0$). 
The Laplacian $\mathcal{L}=[l_{ij}]\in\mathbb{R}^{N\times N}$ is defined by $l_{ii}=\sum_{j\neq i}a_{ij}$ and $l_{ij}=-a_{ij}$ for $i\neq j$.

\begin{assumption} \label{ass:graph}
The communication graph $\mathcal{G}$ is connected.
\end{assumption}

\smallskip 

Each agent $i$ is assigned an actuation subspace $\mathscr{Z}_i\subseteq\operatorname{Im}B_i$, possibly a proper subspace of its available input channel, with $r_i=\dim\mathscr{Z}_i$ and basis matrix $Z_i\in\mathbb{R}^{n\times r_i}$. Set $Z=[Z_1\ \cdots\ Z_N]$. The tuple $\mathscr{Z}=(\mathscr{Z}_1,\ldots,\mathscr{Z}_N)$ is referred to as the prescribed actuation profile. Restricting the actuation of agent~$i$ to $\mathscr{Z}_i$ shrinks the unknown-input channels seen by the other agents, at the expense of actuation authority.



For system~\eqref{eq:plant}, the agent $i$ has direct access only to its local input $u_i$ and measurement $y_i$ and the estimates received from its neighbors, as illustrated in Fig.~\ref{fig:multichannel}. Note that the agent $i$ neither knows the feedback gains nor the control inputs of the other agents. 
The objective is to design, for each agent, a local observer and an estimate-based feedback law using only available information, such that for all initial conditions of the plant and the observers
\begin{equation}
\lim_{t\to\infty}\|x(t)\|=0,\quad
\lim_{t\to\infty}\|x(t)-\hat x_i(t)\|=0,
\quad i\in\mathbf{N},
\label{eq:objective}
\end{equation}
where $\hat x_i$ is the local estimate maintained by agent $i$.

\section{Distributed Output-Feedback Control}\label{sec:dof_control}
\subsection{Geometric Observer-Controller Design}\label{sec:CO_design}
Since $\mathscr Z_i\subseteq\operatorname{Im}B_i$, choose $R_i$ such that $B_iR_i=Z_i$. Each agent then employs the local estimate-based feedback law
\begin{equation}
 u_i=R_iH_i\hat x_i,\qquad
 B_i u_i=Z_iH_i\hat x_i\in\Z_i.
 \label{eq:feedback}
\end{equation}
where $H_i:\X\rightarrow\mathbb R^{r_i}$ denotes the local feedback map of agent $i$, and $r_i=\dim\mathscr Z_i$. The stacked gain $H=\col(H_1,\ldots,H_N)$ belongs to a prescribed nonempty compact set $\Hs$. 

During operation, agents share their local state estimates with neighbors, while their selected feedback gain $H_i$ and actual control inputs $u_i$ remain undisclosed. Consequently, for each agent $i$, the aggregate control action $d_i\coloneqq\sum_{j\ne i} B_j u_j$ acts as an unknown input, whose direction is confined to the subspace $\mathscr D_i\coloneqq\sum_{j\ne i}\Z_j$, i.e., $d_i\in \mathscr D_i$.

To decouple this unknown input while retaining as much state information as possible, consider the family of $(C_i,A)$-invariant subspaces containing $\mathscr D_i$ that admit an output injection rendering the induced quotient map Hurwitz:
\begin{equation*}
    \begin{aligned}
\underline{\mathfrak W}_{g,i}(\mathscr D_i)
    \coloneqq\bigl\{\mathscr W_i:\;&
    \W_i\in\underline{\W}(C_i,A;\mathscr D_i),\ \exists L_i\in\mathbf L(\mathscr W_i)
    \text{  s.t.}\\
    &\kappa((A+L_iC_i)|{\mathscr X/\mathscr W_i}) \subset\mathbb C_g
    \bigr\}.
\end{aligned}
\end{equation*}
Following the infimal-subspace construction in~\cite{zhao2026duio}, define $\mathscr{W}^{*}_{g,i}:=\inf\underline{\mathfrak W}_{g,i}(\mathscr D_i)$. We call $\mathscr{W}^{*}_{g,i}$ the local loss subspace of agent~$i$, as it contains the unknown-input subspace $\mathscr{D}_i$ together with all state directions that agent~$i$ cannot reconstruct with a stable error from $y_i$ alone.

Let $W_i$ and $M_i$ be an orthonormal transformation adapted to the decomposition $\X = \W_{g,i}^* \oplus \mathscr M_i$, where $\mathscr M_i \simeq \X/\W_{g,i}^*$.
For $L_i\in\mathbf{L}(\W_{g,i}^*)$, the matrix representation of the restriction map $(A+L_iC_i)|{\mathscr W_{g,i}^*}$ and the induced map $(A+L_iC_i)|\X/\mathscr W_{g,i}^*$ are given by
\begin{equation*}
\begin{aligned}
&\widetilde{A}_{L_i}\coloneq \mathrm{Mat}((A+L_iC_i)|\W_{g,i}^*)={W_i}^\top (A+L_iC_i) W_{i}\label{eq:tildeAL}\\
&\Bar{A}_{L_i}\coloneq \mathrm{Mat}((A+L_iC_i)|\X/\W_{g,i}^*)=M_i^\top (A+L_iC_i) M_i,
\end{aligned}
\end{equation*}
respectively, and satisfy
\begin{equation*}\label{eq:quotient_design}
    W_i\widetilde{A}_{L_i}=(A+L_iC_i)W_i,\ \Bar{A}_{L_i}M_i^\top=M_i^\top(A+L_iC_i),
\end{equation*}
where $\bar A_{L_i}$ is Hurwitz by construction.

The local observer of agent $i$ is
\begin{equation}
 \begin{split}
 \dot{\hat x}_i={}&A\hat x_i+L_i(C_i\hat x_i-y_i)+B_i u_i\\
 &+\rho W_iW_i^\top\sum_{j=1}^N a_{ij}(\hat x_j-\hat x_i),
 \end{split}
 \label{eq:observer}
\end{equation}
where $\rho>0$ is a common consensus gain. The projection $W_iW_i^{\top}$ confines the consensus correction to the loss subspace $\mathscr{W}^{*}_{g,i}$, where local information is insufficient to observe. Together, \eqref{eq:feedback} and~\eqref{eq:observer} define the distributed output-feedback scheme.

The construction above is stated in the geometric form of~\cite{zhao2026duio}, since $\mathscr{W}^{*}_{g,i}$, the admissible injections, and the stability of the induced quotient map are independent of coordinates. The subsequent analysis, however, uses only the following consequences in the orthonormal bases $W_i$, $M_i$: for all $i\in\mathbf{N}$ and $j\neq i$,
\begin{subequations}\label{eq:geo2mat}
\begin{align}
  & W_iW_i^{\top}+M_iM_i^{\top}=I_n,\qquad M_i^{\top}W_i=0, \label{eq:geo2mat-a}\\
  & M_i^{\top}(A+L_iC_i)=\bar A_{L_i}M_i^{\top},\quad \bar A_{L_i}\ \text{Hurwitz}, \label{eq:geo2mat-b}\\
  & M_i^{\top}Z_j=0. \label{eq:geo2mat-c}
\end{align}
\end{subequations}
Here, \eqref{eq:geo2mat-b} expresses the $(A+L_iC_i)$-invariance of $\mathscr{W}^{*}_{g,i}$ with a stable induced quotient map, and \eqref{eq:geo2mat-c} follows from $\mathscr{Z}_j\subseteq\mathscr{D}_i\subseteq\mathscr{W}^{*}_{g,i}$. Hence, from Section~\ref{sec:analysis} onward, the analysis is carried out in matrix form. Any other choice of orthonormal bases is related to this one by orthogonal transformations, which leave all spectra unchanged, and the choice is therefore without loss of generality.

\subsection{Closed-loop Stability Analysis}\label{sec:analysis}
Let $e_i=x-\hat x_i$, $e=\col_{i\in\mathbf{N}}(e_i)$, $G_i=Z_iH_i$, and $G=ZH=\sum_{i=1}^N G_i$. Substituting \eqref{eq:feedback} and \eqref{eq:observer} into the plant and error equations gives
\begin{equation*}
 \begin{split}
 \dot x=&(A+G)x-\sum_{i=1}^N G_i e_i,\\
 \dot e_i\!=&(A\!+\!L_iC_i)e_i\!+\!\sum_{\substack{j\ne i\\j\in\mathbf{N}}}\!G_j(x\!-\!e_j)\!-\!\rho W_iW_i^\top\!\sum_{j=1}^N \!a_{ij}(e_i\!-\!e_j).
 \end{split}
 \label{eq:error}
\end{equation*}
With $e=\col(e_1,\ldots,e_N)$, the complete closed loop is
\begin{equation}
\begin{bmatrix}
    \dot{x}\\\dot{e}
\end{bmatrix}=\begin{bmatrix}
A+G&-[G_1\ \cdots\ G_N]\\
\col(G-G_i)_{i=1}^N&E(H,\rho)
\end{bmatrix}\begin{bmatrix}
    x\\e
\end{bmatrix}
\label{eq:acl}
\end{equation}
where the blocks of $E(H,\rho)=[E_{ij}]_{i,j\in\mathbf{N}}$ are given by
\begin{equation*}
E_{ij}=
\begin{cases}
A+L_iC_i-\rho\bigl(\sum_{k=1}^N a_{ik}\bigr)W_iW_i^\top,&i=j,\\
-G_j+\rho a_{ij}W_iW_i^\top,&i\ne j.
\end{cases}
\label{eq:aclblocks}
\end{equation*}
Let $\mathscr{W}_{\oplus}=\mathscr{W}^{*}_{g,1}\times\cdots\times\mathscr{W}^{*}_{g,N}$
collect the stacked errors. Its common-error subspace is $\W_{\mathrm c} \coloneqq \left\{ \mathbf 1_N\otimes s:s\in\Ss \right\}$, where $\Ss=\bigcap_{i=1}^N\W_{g,i}^*$. Consequently, $\W_\oplus$ admits the orthogonal
direct-sum decomposition
\begin{equation*}
\W_\oplus
=\W_{\mathrm c}\oplus
(\W_\oplus\cap\W_{\mathrm c}^{\perp}).
\footnote{In~\cite{wonham1985}, $\mathscr{W}^{\perp}$ denotes the
annihilator of $\mathscr{W}$, a subspace of the dual space. Here, with the
standard bases and the Euclidean inner product fixed, the dual space is
identified with the space itself, and
$\mathscr{W}^{\perp}=\{z: z^{\top}w=0,\ \forall w\in\mathscr{W}\}$ denotes the
orthogonal complement. Hence here $\mathscr{W}_{\mathrm{c}}^{\perp}\subseteq\mathbb{R}^{Nn}$.
This choice depends on a certain coordinate, which is why the analysis below
is carried out in matrix form.}
\label{eq:loss_decomposition}
\end{equation*}
Let $U\in\mathbb R^{n\times d}$ be an orthonormal basis of $\Ss$, where $n_s=\dim\Ss$. Then, we define $\overline U=\mathbf 1_N\otimes U$, and choose an orthonormal basis $T=\col(T_1,\ldots,T_N)$ of $\W_\oplus\cap\W_{\mathrm c}^{\perp}$. Together with $M=\operatorname{diag}(M_1,\ldots,M_N)$, the matrix
\[
\left[
\frac{\overline U}{\sqrt N}\quad T\quad M
\right]
\]
is orthogonal. Consequently, the stacked estimation error admits the unique decomposition
\begin{equation}
e=\overline Uq+T\xi+M\zeta,
\label{eq:direct_decomp}
\end{equation}
where $q=\frac1N\overline U^\top e$, $\xi=T^\top e$, $\zeta=M^\top e$. Here $\zeta=\col(\zeta_1,\ldots,\zeta_N)$ with $\zeta_i=M_i^\top e_i$. In particular, $e_i=Uq+T_i\xi+M_i\zeta_i$.

Let $\widetilde x=\operatorname{col}(x,q)$. Since $G_j=Z_jH_j$, \eqref{eq:geo2mat-c} gives $M_i^{\top}G_j=0$ for $j\neq i$.
Using this together with \eqref{eq:geo2mat-a}--\eqref{eq:geo2mat-b} substituting \eqref{eq:direct_decomp} into \eqref{eq:acl} yields
\begin{equation}
\begin{bmatrix}
\dot{\widetilde x}\\
\dot\xi\\
\dot\zeta
\end{bmatrix}
=\begin{bmatrix}
A_{\mathrm{com}}(H,L)&B_s(H)&\star\\
C_s(H)&D_s(H)-\rho\Lambda&\star\\
0&0&A_\zeta
\end{bmatrix}
\begin{bmatrix}
\widetilde x\\
\xi\\
\zeta
\end{bmatrix},
\label{eq:direct_closed_loop}
\end{equation}
where $\star$ are blocks irrelevant to the spectrum,
\begin{equation*}
\begin{aligned}
&A_{\mathrm{com}}(H,L)
=\\
&\resizebox{0.98\columnwidth}{!}{$\displaystyle
\begin{bmatrix}
A+G & -GU\\
\frac{N-1}{N}U^\top G &
U^\top AU+\frac1N\sum_{i=1}^N U^\top L_iC_iU
-\frac{N-1}{N}U^\top GU
\end{bmatrix}
$},\\
&B_s(H)
=
\begin{bmatrix}
-[G_1\ \cdots\ G_N]T\\
N^{-1}\overline U^\top E(H,0)T
\end{bmatrix},\ L=[L_1\ \cdots\ L_N],\\
&C_s(H)
=T^\top
\begin{bmatrix}
\operatorname{col}(G-G_i)_{i=1}^N&
E(H,0)\overline U
\end{bmatrix},\\
&D_s(H)=T^\top E(H,0)T,\ 
\Lambda=T^\top(\mathcal L\otimes I_n)T,
\end{aligned}
\label{eq:direct_blocks}
\end{equation*}
and $A_\zeta= \operatorname{diag} (\Bar{A}_{L_1},\cdots,\Bar{A}_{L_N})$. Here, $E(H,0)$ denotes $E(H,\rho)$ evaluated at $\rho=0$.

To provide a sufficient condition for the exponential convergence of the complete closed loop, we impose the following assumption.

\begin{assumption}\label{ass:geometry}
$\kappa(A_{\mathrm{com}}(H,L))\subset\mathbb C_g,\  \forall H\in\Hs$.
\end{assumption}

As shown in \eqref{eq:direct_closed_loop}, the consensus term acts directly only on the disagreement coordinates \(\xi\) and vanishes on the common-error subspace \(\W_c\). The stability requirement on the coupled plant and common-error dynamics is therefore addressed through the feedback gains \(H_i\) and output injections \(L_i\), allowing a nontrivial intersection \(\Ss\ne 0\) of the local loss subspaces. When \(\Ss=0\), the common-error coordinates disappear, and Assumption~\ref{ass:geometry} reduces to \(A+ZH\) being Hurwitz for every \(H\in\Hs\). A sufficient synthesis procedure to enforce Assumption~\ref{ass:geometry} will be introduced in Section~\ref{sec:synthesis}.

\vspace{0.5em}

\begin{theorem}\label{thm:main}
Consider system~\eqref{eq:plant} with the feedback~\eqref{eq:feedback} and observer~\eqref{eq:observer}. Under Assumptions~\ref{ass:graph}--\ref{ass:geometry}, for every fixed $H\in\Hs$, the exponential convergence in \eqref{eq:objective} is guaranteed for all initial conditions if $T$ has at least one column and 
\begin{equation}
\rho>\max\left\{0,\,\sup_{H\in\Hs}\inf_{\tau>0}\lambda_{\max}\left(\Lambda^{-1/2}\Phi_H(\tau)\Lambda^{-1/2}\right)\right\},
\label{eq:rho_bound}
\end{equation}
where
\begin{align*}
\Phi_H(\tau)&=\frac{\operatorname{He}(D_s(H))}{2}+\frac{K_H(\tau)^\top K_H(\tau)}{2\tau},\\
K_H(\tau)&=\tau P_HB_s(H)+C_s(H)^\top,
\end{align*}
and $P_H=P_H^\top\succ0$ is the unique solution of
\begin{equation}
\operatorname{He}\bigl(P_HA_{\mathrm{com}}(H,L)\bigr)=-I_{n+n_s}.
\label{eq:lyap_PH}
\end{equation}
If $T$ has no columns, the conclusion holds for every $\rho>0$.
\end{theorem}
\begin{proof}
According to Assumption~\ref{ass:geometry}, $A_{\mathrm{com}}(H,L)$ is Hurwitz for every $H\in\Hs$. If $T$ has no columns, the disagreement subspace $\W_\oplus\cap\W_c^\perp$ is trivial, the matrix in~\eqref{eq:direct_closed_loop} is block triangular with Hurwitz diagonal blocks $A_{\mathrm{com}}(H,L)$ and $A_\zeta$. Hence, the conclusion holds for every fixed $H\in\Hs$ and $\rho>0$.

Suppose that the disagreement subspace $\W_\oplus\cap\W_c^\perp$ is nontrivial. Since the communication graph is undirected and connected, we have
\[
\mathcal L\otimes I_n\succeq0,
\qquad
\ker(\mathcal L\otimes I_n)\cap\W_\oplus
=\W_{\mathrm c}.
\]
If $\xi^\top\Lambda\xi=0$, positive semidefiniteness implies $T\xi\in\ker(\mathcal L\otimes I_n)$. By the construction of $T$, $T\xi\in\W_\oplus\cap\W_{\mathrm c}^{\perp}$. Consequently, $T\xi\in
\W_{\mathrm c}\cap\W_{\mathrm c}^{\perp}
=0$. Since $T^\top T=I$, we obtain $\xi=0$.
Thus $\Lambda\succ0$ holds.

For each fixed $H\in\Hs$, the Hurwitz property of $A_{\mathrm{com}}(H,L)$ guarantees a unique positive definite solution $P_H$ of \eqref{eq:lyap_PH}.
Consider the homogeneous $(\widetilde x,\xi)$ subsystem of~\eqref{eq:direct_closed_loop}, obtained by setting $\zeta=0$, and let
\[
V=\tau\widetilde x^\top P_H\widetilde x
+\|\xi\|^2,
\qquad \tau>0.
\]
Using~\eqref{eq:lyap_PH}, its derivative is
\begin{equation*}
\dot V
=-\tau\|\widetilde x\|^2
+2\widetilde x^\top K_H(\tau)\xi+\xi^\top
\bigl(\operatorname{He}(D_s(H))-2\rho\Lambda\bigr)
\xi.
\label{eq:lyap_derivative}
\end{equation*}
Thus $\dot V$ is negative definite if and only if
\[
\begin{bmatrix}
-\tau I_{n+n_s}&K_H(\tau)\\
K_H(\tau)^\top&
\operatorname{He}(D_s(H))-2\rho\Lambda
\end{bmatrix}\prec0.
\]
Since $-\tau I_{n+n_s}\prec0$, the Schur complement gives the equivalent condition
\[
\operatorname{He}(D_s(H))-2\rho\Lambda +\tau^{-1}K_H(\tau)^\top K_H(\tau)\prec0.
\]
By the definition of $\Phi_H(\tau)$, this is equivalent to
\[
\rho\Lambda\succ\Phi_H(\tau).
\]
As $\Lambda\succ0$, it follows that $\dot V$ is negative definite precisely when
\begin{equation}
\rho>
\lambda_{\max}\left(
\Lambda^{-1/2}\Phi_H(\tau)\Lambda^{-1/2}
\right).
\label{eq:schur_rho}
\end{equation}
Choose any fixed $\rho$ satisfying \eqref{eq:rho_bound}. For every fixed
$H\in\Hs$,
\[
\inf_{\tau>0}\lambda_{\max}\left(\Lambda^{-1/2}\Phi_H(\tau)\Lambda^{-1/2}\right)<\rho.
\]
By the definition of the infimum, there exists a finite $\tau_H>0$ such that \eqref{eq:schur_rho} holds. For this choice, $V$ is positive definite and $\dot V$ is negative definite.
Therefore, the homogeneous $(\widetilde x,\xi)$ subsystem is exponentially stable for each fixed $H\in\Hs$ and each $\rho$ satisfying~\eqref{eq:rho_bound}. Since $A_\zeta$ is Hurwitz, the block triangular structure of~\eqref{eq:direct_closed_loop} implies exponential stability of the complete transformed system. Since~\eqref{eq:direct_decomp} is an invertible coordinate transformation, $x(t)$ and $e_i(t)=x(t)-\hat x_i(t)$ converge exponentially to zero for all initial conditions, proving~\eqref{eq:objective}, which completes the proof.
\end{proof}
\begin{remark}
The distributed unknown-input observer designs in \cite{zhao:ecc25,zhao2026duio}, require the joint geometric condition $\Ss=\bigcap_{i=1}^N\W_{g,i}^*=0$, which ensures that all loss-space error directions can be suppressed by consensus. The present framework permits a nontrivial intersection under Assumption~\ref{ass:geometry}. This relaxation concerns joint output-feedback stabilization, where remote control inputs are generated by the prescribed feedback laws rather than arbitrary exogenous signals.
\end{remark}

\section{Observer--Controller Synthesis}
\label{sec:synthesis}
To enforce the design-dependent Hurwitz condition in Assumption~\ref{ass:geometry}, this section develops a sufficient synthesis procedure over a prescribed feedback polytope. We parameterize the local output injections to preserve quotient stability and jointly search for nominal feedback and injection gains with a common Lyapunov certificate via alternating LMIs. A subsequent vertex-based SDP certifies a common consensus gain, guaranteeing exponential closed-loop convergence for every fixed feedback selection in the certified polytope.

\subsection{Parameterization of the Output Injections}
For each $i\in\mathbf N$, choose an $L_i^0\in\mathbf L(\W_{g,i}^*)$ \cite[Algorithm~1-3]{zhao2026duio}, which satisfies
\[
M_i^\top(A+L_i^0C_i)W_i=0,
\qquad \bar A_{L_i^0}\text{ is Hurwitz}.
\]
Introduce a free design matrix $F_i\in\R^{n_s\times p_i}$ and set
\begin{equation}
 L_i=(I_n-UU^\top)L_i^0+NUF_i.
 \label{eq:syn_injection}
\end{equation}
Since $\Ss=\bigcap_{i=1}^N\W_{g,i}^*\subseteq\W_{g,i}^*$, we have $M_i^\top U=0$ and therefore
\begin{equation*}
M_i^\top(A+L_iC_i)=M_i^\top(A+L_i^0C_i),\ 
U^\top L_i=NF_i.
\label{eq:syn_injection_properties}
\end{equation*}
Thus conditioned invariance and the Hurwitz matrices $\bar A_{L_i}=\bar A_{L_i^0}$ are preserved for every $F_i$. Consequently, $A_\zeta$ remains Hurwitz.

Let $F=[F_1\ \cdots\ F_N]$, $C=\col(C_1,\ldots,C_N)$, and $C_U=CU$.
Then
\begin{equation}
 \frac1N\sum_{i=1}^N U^\top L_iC_iU
 =\sum_{i=1}^N F_iC_iU=FC_U.
 \label{eq:syn_common_injection}
\end{equation}
Substituting~\eqref{eq:syn_common_injection} into the previously defined $A_{\mathrm{com}}(H,L)$ yields the parameterized matrix
\begin{equation}
\begin{aligned}
&A_{\mathrm{com}}(H,F)=\\
&\begin{bmatrix}
A+ZH & -ZHU\\
\frac{N-1}{N}U^\top ZH &
U^\top AU+FC_U-\frac{N-1}{N}U^\top ZHU
\end{bmatrix},
\end{aligned}\label{eq:syn_acom}
\end{equation}
which is jointly affine in $H$ and $F$. The objective is therefore to find one injection gain $F$ that stabilizes these reduced dynamics throughout the admissible feedback family
\begin{equation*}
 \exists F\quad\forall H\in\Hs:\quad
 A_{\mathrm{com}}(H,F)\text{ is Hurwitz}.
 \label{eq:syn_shared_injection}
\end{equation*}

\subsection{Joint Design of a Feedback Family}
To accommodate undisclosed local feedback selections, we seek a nominal gain $H_0$ that stabilizes a prescribed family of gain variations. Specifically, let $\Delta^{(1)},\ldots,\Delta^{(r)}$ be prescribed gain offsets satisfying $0\in\operatorname{co}\{\Delta^{(1)},\ldots,\Delta^{(r)}\}$. The admissible feedback family is parameterized as
\begin{equation}
\begin{aligned}
 H^{(k)}&=H_0+\Delta^{(k)},\qquad k=1,\ldots,r,\\
 \Hs(H_0)&=\operatorname{co}\{H^{(1)},\ldots,H^{(r)}\}.
\end{aligned}
\label{eq:syn_family}
\end{equation}
The offsets determine the shape and size of the feedback polytope, while the nominal gain $H_0$ determines its position. After synthesis, $H_0$ and the family $\Hs(H_0)$ are fixed. Choosing the offset polytope as a Cartesian product of local admissible sets permits independent local selections whose stacked gain belongs to $\Hs(H_0)$. Each selected gain is held constant during operation.

Our objective is to jointly design $H_0$ and $F$ such that $A_{\mathrm{com}}(H,F)$ is Hurwitz for every $H\in\Hs(H_0)$. To obtain a tractable stability certificate, we require a common quadratic Lyapunov function over the entire feedback polytope. Specifically, seek $H_0$, $F$, and a symmetric matrix $P\in\R^{(n+n_s)\times(n+n_s)}$ satisfying
\begin{equation}
\begin{aligned}
P&\succ0,\\
\operatorname{He}\bigl(
PA_{\mathrm{com}}(H^{(k)},F)
\bigr)&\prec0,\quad k=1,\ldots,r.
\end{aligned}
\label{eq:syn_joint_bmi}
\end{equation}
Since $A_{\mathrm{com}}(H,F)$ is affine in $H$, the same Lyapunov inequality holds for every $H\in\Hs(H_0)$ by convexity. Hence, \eqref{eq:syn_joint_bmi} guarantees Assumption~\ref{ass:geometry} for the entire feedback family. These conditions are bilinear in $P$ and $(H_0,F)$, but become LMIs when either block is fixed. A common quadratic certificate is sufficient, not necessary, for $A_{\mathrm{com}}(H,F)$ to be Hurwitz throughout the family.

\subsection{Alternating LMI Algorithm}
To search for a solution of~\eqref{eq:syn_joint_bmi}, we alternate between the gain variables $(H_0,F)$ and the Lyapunov matrix $P$, following the iterative linear matrix inequality (LMI) approach to output-feedback synthesis~\cite{leibfritz2001output}. Prescribe $h_{\max},f_{\max}>0$ and impose
\begin{equation}
 \|H_0\|_2\le h_{\max},\qquad \|F\|_2\le f_{\max}.
 \label{eq:syn_gain_bounds}
\end{equation}
These spectral-norm bounds admit LMI representations and keep the gain search bounded. To remove the scaling freedom of $P$ and limit ill-conditioning, choose $0<\delta<1$ and require
\begin{equation}
 P\succeq\delta I_{n+n_s},\qquad \operatorname{tr}P=n+n_s.
 \label{eq:syn_normalization}
\end{equation}
The gain bounds and the prescribed lower bound $\delta$ are numerical design restrictions and may exclude otherwise feasible certificates.

Introduce a scalar slack $\sigma\ge-1$, where $-1$ is an arbitrary lower bound on the auxiliary objective, and impose
\begin{equation}
 \operatorname{He}\!\left(
 PA_{\mathrm{com}}(H_0+\Delta^{(k)},F)\right)
 \preceq\sigma I_{n+n_s},
 \quad k=1,\ldots,r.
 \label{eq:syn_slack}
\end{equation}
The slack bounds the largest eigenvalue of the Lyapunov residuals over all vertices. A negative $\sigma$ certifies~\eqref{eq:syn_joint_bmi}.

Initialize $P^{(0)}=I_{n+n_s}$. At iteration $\ell$, perform the following two updates.

\emph{Gain update:} With $P^{(\ell-1)}$ fixed, solve
\begin{equation}
\begin{aligned}
\underset{H_0,F,\sigma}{\operatorname{minimize}}
&\quad \sigma\\
\text{subject to}
&\quad \eqref{eq:syn_gain_bounds},\quad \sigma\ge-1,\\
&\quad \operatorname{He}\!\left(
P^{(\ell-1)}
A_{\mathrm{com}}(H_0+\Delta^{(k)},F)\right)\\
&\qquad\preceq\sigma I_{n+n_s},
\quad k=1,\ldots,r.
\end{aligned}
\label{eq:syn_gain_update}
\end{equation}
Denote the solution by $(H_0^{(\ell)},F^{(\ell)},\sigma_{\mathrm g}^{(\ell)})$.

\emph{Certificate update:} With $(H_0^{(\ell)},F^{(\ell)})$ fixed, solve
\begin{equation}
\begin{aligned}
\underset{P=P^\top,\sigma}{\operatorname{minimize}}
&\quad \sigma\\
\text{subject to}
&\quad \eqref{eq:syn_normalization},\quad \sigma\ge-1,\\
&\quad \operatorname{He}\!\left(
PA_{\mathrm{com}}
(H_0^{(\ell)}+\Delta^{(k)},F^{(\ell)})\right)\\
&\qquad\preceq\sigma I_{n+n_s},
\quad k=1,\ldots,r.
\end{aligned}
\label{eq:syn_certificate_update}
\end{equation}
Denote the solution by $(P^{(\ell)},\sigma^{(\ell)})$.

Each subproblem has a linear objective and LMI constraints, and is therefore an SDP~\cite{vandenberghe1996sdp}. Algorithm~\ref{alg:synthesis} summarizes the procedure.

\begin{algorithm}[t]
\small
\caption{Alternating LMI synthesis}
\label{alg:synthesis}
\LinesNumbered
\KwIn{Fixed model and geometry; quotient-stabilizing
baseline friends $L_i^0$; offsets $\Delta^{(k)}$;
bounds $h_{\max},f_{\max}>0$;
$0<\delta,\varepsilon<1$;
stagnation tolerance $\varepsilon_s\ge0$;
iteration limit $K_{\max}$.}

$P^{(0)}\leftarrow I_{n+n_s}$\;

\For{$\ell=1,\ldots,K_{\max}$}{

  Solve~\eqref{eq:syn_gain_update} to obtain
  $(H_0^{(\ell)},F^{(\ell)},\sigma_{\mathrm g}^{(\ell)})$\;

  \If{no verified feasible gain update is obtained}{
    \AlgReturn{inconclusive}\;
  }

  Solve~\eqref{eq:syn_certificate_update} to obtain
  $(P^{(\ell)},\sigma^{(\ell)})$\;

  \If{no verified feasible certificate update is obtained}{
    \AlgReturn{inconclusive}\;
  }

  \If{$\sigma^{(\ell)}\le-\varepsilon$ and
  $(H_0^{(\ell)},F^{(\ell)},P^{(\ell)})$
  satisfies~\eqref{eq:syn_joint_bmi}
  upon independent verification}{
    \AlgReturn{$(H_0^{(\ell)},F^{(\ell)},P^{(\ell)})$}\;
  }

  \If{$\ell>1$ and
  $|\sigma^{(\ell)}-\sigma^{(\ell-1)}|
  \le\varepsilon_s$}{
    \AlgBreak\;
  }
}

\AlgReturn{inconclusive}\;

\KwOut{A verified certificate $(H_0,F,P)$
or an inconclusive result.}
\end{algorithm}

The slack formulation does not require stabilizing initial gains. Under exact alternating updates, the optimized slack is nonincreasing, but convergence to a negative value is not guaranteed. Only solutions satisfying~\eqref{eq:syn_joint_bmi} are accepted. Otherwise, the search is inconclusive. When $d=0$, $F$ and its associated constraints are omitted.

\subsection{Consensus-Gain Certification}

Fix a certificate $(H_0,F,P)$ satisfying~\eqref{eq:syn_joint_bmi} and recover $L_i$ from~\eqref{eq:syn_injection}. It remains to select a consensus gain $\rho$ that stabilizes the interconnection between the plant--common-error and disagreement dynamics. Rather than directly evaluating~\eqref{eq:rho_bound}, which involves gain-dependent Lyapunov solutions $P_H$, we exploit the common certificate $P$ to formulate a finite set of vertex LMIs.

If $T$ has no columns, the disagreement dynamics are absent and any $\rho>0$ suffices. Suppose $T$ has at least one column, so that $\Lambda\succ0$ under Assumption~\ref{ass:graph}. Consider the Lyapunov function $V=\tau\widetilde x^\top P\widetilde x+\|\xi\|^2, \ \tau>0$, for the homogeneous $(\widetilde x,\xi)$ subsystem. Let superscript $(k)$ denote evaluation at $H^{(k)}$ with the fixed injections $L_i$. Negative definiteness of $\dot V$ at each vertex is equivalent to
\begin{equation}
\begin{aligned}
\mathcal M_k(\tau,\rho)\coloneq
\begin{bmatrix}
\tau\operatorname{He}(PA_{\mathrm{com}}^{(k)})
&\tau PB_s^{(k)}+(C_s^{(k)})^\top\\
\tau {B_s^{(k)}}^\top P + C_s^{(k)}&\operatorname{He}(D_s^{(k)})-2\rho\Lambda
\end{bmatrix}
\prec0,
\end{aligned}
\label{eq:syn_rho_lmi}
\end{equation}
for $k=1,\ldots,r$. Since $\mathcal M_k$ is affine in $(\tau,\rho)$, the consensus-gain threshold is characterized by
\begin{equation}
\begin{aligned}
\rho_{\mathrm{SDP}}
\coloneq\inf_{\tau,\rho}\quad&\rho\\
\text{subject to}\quad
&\tau>0,\quad\rho\ge0,\\
&\mathcal M_k(\tau,\rho)\prec0,
\quad k=1,\ldots,r.
\end{aligned}
\label{eq:syn_rho_sdp}
\end{equation}

Since $\operatorname{He}(PA_{\mathrm{com}}^{(k)})\prec0$ and $\Lambda\succ0$, the Schur complement guarantees feasibility of~\eqref{eq:syn_rho_lmi} for sufficiently large $\rho$, implying $\rho_{\mathrm{SDP}}<\infty$. Furthermore, $\mathcal M_k(\tau,\rho+\eta)\preceq\mathcal M_k(\tau,\rho),\  \forall\eta\ge0$. Thus, every $\rho>\rho_{\mathrm{SDP}}$ admits a common $\tau>0$ satisfying~\eqref{eq:syn_rho_lmi} at all vertices.

For any $H\in\Hs(H_0)$, there exist $\alpha_k\ge0$ with $\sum_{k=1}^r\alpha_k=1$ such that $H=\sum_{k=1}^r\alpha_kH^{(k)}$. By affine dependence on $H$,
\[
\mathcal M(H,\tau,\rho)
=\sum_{k=1}^r\alpha_k\mathcal M_k(\tau,\rho)
\prec0.
\]
Together with the Hurwitz block $A_\zeta$ and the block triangular structure in~\eqref{eq:direct_closed_loop}, this guarantees exponential convergence in~\eqref{eq:objective} for every fixed $H\in\Hs(H_0)$.



\section{Simulation Results}\label{sec:example} Consider three agents connected by the undirected path $1$--$2$--$3$. Let $b_j$ denote the $j$th column of $I_5$, and set
\begin{equation*}
A=\left[\begin{smallmatrix}
 0.8&0.7&0&0&0.1\\
 -0.2&-0.5&0&0&0.2\\
 0&0&-1&1.5&0\\
 0&0&1.5&-1&0\\
 0&0&0&0&0.3
\end{smallmatrix}\right].
\label{eq:example_A}
\end{equation*}
The input and measurement matrices are given by
\begin{equation*}
\begin{aligned}
 B_1=Z_1&=[b_1\ b_5],& C_1&=[b_1\ b_5]^\top,\\
 B_2=Z_2&=b_2,& C_2&=b_3^\top,\\
 B_3=Z_3&=b_3,& C_3&=(b_1+b_5)^\top.
\end{aligned}
\label{eq:example_channels}
\end{equation*}
The pairs $(A,Z)$ and $(C,A)$ are controllable and observable, respectively, whereas the local observability ranks are $3,2,3$, and none of the local pairs is detectable. For $\bar B_i=[Z_j]_{j\ne i}$, the ranks of $\bar B_i$ are $2,3,3$, while those of $C_i\bar B_i$ are $0,1,1$. Thus, the local unknown inputs cannot be estimated by individual local information \cite{zhao2026duio}.

The geometric construction yields
\begin{equation*}
 W_1=\left[\begin{smallmatrix}
1 & 0 & 0 & 0 \\ 
0 & 1 & 0 & 0 \\ 
0 & 0 & 1 & 0 \\ 
0 & 0 & 0 & 1 \\ 
0 & 0 & 0 & 0
\end{smallmatrix}\right], 
W_2=\left[\begin{smallmatrix}
1 & 0 & 0 & 0 \\ 
0 & 1 & 0 & 0 \\ 
0 & 0 & 1 & 0 \\ 
0 & 0 & 0 & 0 \\ 
0 & 0 & 0 & 1
\end{smallmatrix}\right], 
W_3=\left[\begin{smallmatrix}
1 & 0 & 0 & 0 \\ 
0 & 1 & 0 & 0 \\ 
0 & 0 & 0.7071 & 0 \\ 
0 & 0 & 0.7071 & 0 \\ 
0 & 0 & 0 & 1
\end{smallmatrix}\right].
\label{eq:example_loss_spaces}
\end{equation*}
Consequently, $\Ss\simeq\operatorname{span}\{b_1,b_2\}$, $U=[b_1\ b_2]$, and $(n_s,\dim\xi,\dim\zeta)=(2,10,3)$. Choosing $L_i^0$ as follows
 \begin{equation*}
 L_1^0=\left[\begin{smallmatrix}
0 & 0 \\ 
0 & 0 \\ 
0 & 0 \\ 
0 & 0 \\ 
0 & -2.3
\end{smallmatrix}\right], 
L_2^0=\left[\begin{smallmatrix}
0 \\ 
0 \\ 
0 \\ 
-1.5 \\ 
0
\end{smallmatrix}\right], 
L_3^0=\left[\begin{smallmatrix}
0 \\ 
0 \\ 
0 \\ 
0 \\ 
0
\end{smallmatrix}\right].
\end{equation*}
Moreover, $M_1=b_5$, $M_2=b_4$, and $M_3=(b_3-b_4)/\sqrt2$, this gives $A_\zeta=\operatorname{diag}(-2,-1,-2.5)$ and $\lambda_{\min}(\Lambda)=0.1771$.
Allow independent fixed gain selections
\begin{equation*}
 H_i=H_{0i}+\theta_iD_i,\qquad
 |\theta_i|\le\epsilon_H,\qquad
 D_i=\widehat D_i/\|\widehat D_i\|_2,
 \label{eq:example_gain_family}
\end{equation*}
where
\begin{equation*}
\widehat D_1^{\top}=\left[\begin{smallmatrix}
1&0.1\\
0.3&-0.1\\
-0.2&0.2\\
0.1&0\\
0.2&1
\end{smallmatrix}\right],\
\widehat D_2^{\top}=\left[\begin{smallmatrix}0.2\\1\\0.3\\-0.1\\0.1\end{smallmatrix}\right],\
\widehat D_3^{\top}=\left[\begin{smallmatrix}0.1\\-0.2\\1\\0.3\\-0.1\end{smallmatrix}\right].
\end{equation*}
The nominal design uses $\epsilon_H=0.1$, giving eight vertices. These offsets vary several gain entries but remain restricted to one prescribed matrix direction per agent.

Algorithm~\ref{alg:synthesis} is implemented in YALMIP/MOSEK with $h_{\max}=f_{\max}=5$, $\delta=0.05$, $\varepsilon=10^{-4}$, $\varepsilon_s=10^{-7}$, and $K_{\max}=50$. The synthesized gains are reported in the stacked form $H_0=\operatorname{col}(H_{01},H_{02},H_{03})$ and $F=[F_1\ F_2\ F_3]$
\begin{align*}
    H_0&=\left[\begin{smallmatrix}
-3.7484 & -0.2541 & 0.0082 & -0.0137 & 0.5332 \\ 
-0.0557 & -0.0882 & -0.0116 & 0.0207 & -2.8796 \\ 
-0.2863 & -0.0187 & -0.0002 & 0.0003 & -0.1208 \\ 
-0.0013 & -0.0190 & -4.3990 & -2.3767 & -0.0008
\end{smallmatrix}\right],\\
F&=\left[\begin{smallmatrix}
-3.5064 & 0.0000 & 0.0000 & -3.5059 \\ 
-0.4527 & 0.0000 & 0.0000 & -0.4567
\end{smallmatrix}\right].
\end{align*}
The injections $L_i$ are recovered from~\eqref{eq:syn_injection}.

With this design fixed, solve~\eqref{eq:syn_rho_sdp} using $\tau\ge10^{-7}$ and $\mathcal M_k(\tau,\rho)\preceq-10^{-6}I_{17}$. Solving~\eqref{eq:syn_rho_sdp} yields $\rho_{\mathrm{SDP}}=14.9109$ with $\tau=5.6916$. We select $\rho_{\mathrm{use}}=15.1$, for which the maximum eigenvalue of the vertex LMIs is $-0.0735$, certifying exponential closed-loop stability. For the closed-loop simulation, choose $(\theta_1,\theta_2,\theta_3)=(0.06,-0.04,0.08)$ and $x(0)=(5,-2.5,3,4,-3)^\top$. 

\begin{figure}[htp]
    \centering

    \begin{subfigure}{0.49\columnwidth}
        \centering
        \includegraphics[width=\linewidth]{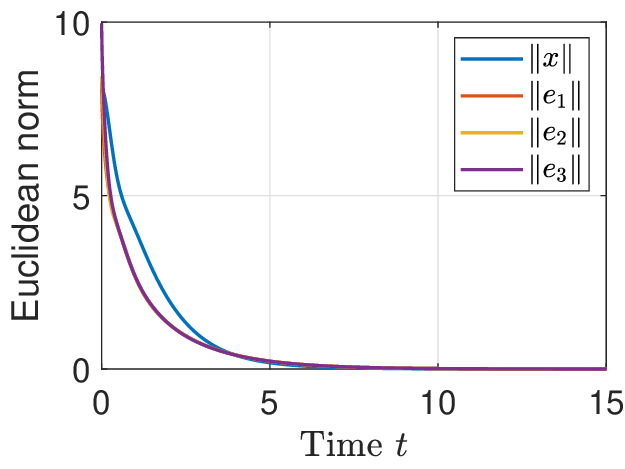}\\[-1.2ex]
        \caption{}
        \label{fig:norm_error_x}
    \end{subfigure}
    \begin{subfigure}{0.49\columnwidth}
        \centering
        \includegraphics[width=\linewidth]{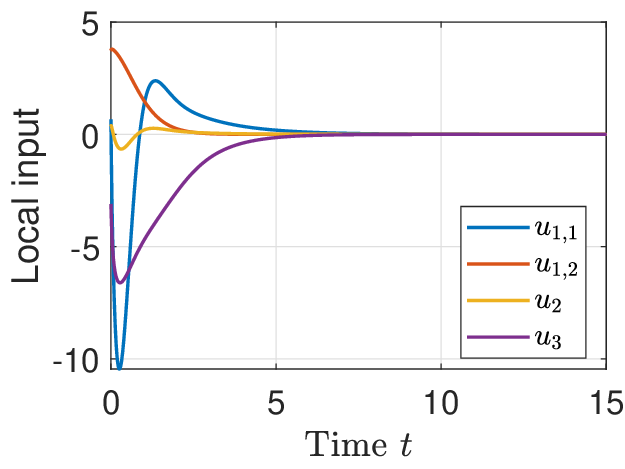}\\[-1.2ex]
        \caption{}
        \label{fig:local_inputs}
    \end{subfigure}
    \vspace{-0.55cm}
    \caption{(a) Norm of local estimation errors and system state. (b) Local input signals of each agent.}
    \label{fig:responses_inputs}
\end{figure}

Fig.~\ref{fig:norm_error_x} shows the convergence of the state and local estimation-error norms, while Fig.~\ref{fig:local_inputs} depicts the corresponding control inputs.
\begin{table}[htp]
\centering
\caption{Feedback-range comparison with fixed $H_0,F,P$}\vspace{-1ex}
\label{tab:example_sensitivity}
\begin{tabular}{ccc}
\hline
$\epsilon_H$ & $\rho_{\mathrm{SDP}}$ & $\rho_{\mathrm{use}}$\\
\hline
0.05 & 14.7482 & 14.9\\
0.10 & 14.9109 & 15.1\\
0.15 & 15.0818 & 15.3\\
0.20 & 15.2662 & 15.5\\
\hline
\end{tabular}
\end{table}
To assess the effect of the admissible feedback range, we vary $\epsilon_H$ while keeping $(H_0,F,P)$ fixed and recomputing $\tau$ and $\rho$ for each case. 
As shown in Table~\ref{tab:example_sensitivity}, the certified consensus-gain threshold increases with $\epsilon_H$, while all four feedback families remain certified.

\section{Conclusions}\label{sec:conclusion}
This paper developed a geometric framework for distributed output-feedback stabilization of linear multichannel systems without exchanging local inputs or feedback gains. The error decomposition isolates stable quotient dynamics while retaining the plant--common-error coupling, thereby accommodating nontrivial intersections of local loss subspaces. A reduced Hurwitz condition and a consensus-gain bound derived from a matrix inequality guarantee exponential closed-loop convergence for every fixed admissible feedback selection. For polytopic feedback families, quotient-preserving output injections and alternating LMIs provide a common quadratic stability certificate, followed by SDP-based consensus-gain certification. Future work will extend the framework to general local control strategies, including nonlinear, optimization-based, and learning-based policies. While the geometric quotient-error decoupling is preserved when their inputs remain within the prescribed actuation subspaces, establishing corresponding closed-loop stability guarantees requires further investigation.


\end{document}